\documentclass[mathleft
]{an}
\usepackage{graphicx}
\usepackage{amsmath}
\usepackage{amssymb}
\usepackage{latexsym}
\usepackage{times}
\usepackage{color}
\usepackage{natbib}
\begin{document}

\Pagespan{789}{}
\Yearpublication{2006}%
\Yearsubmission{2005}%
\Month{11}%
\Volume{999}%
\Issue{88}%

\title{The stability of strong waves and its implications for pulsar wind shocks}

\author{Iwona Mochol\thanks{Corresponding author:
  \email{iwona.mochol@mpi-hd.mpg.de}\newline}
\and  John G. Kirk
}
\titlerunning{Strong waves and their implications for shocks}
\authorrunning{I. Mochol \& J.~G. Kirk}
\institute{
Max-Planck-Institut f\"ur Kernphysik, Saupfercheckweg 1, 
69117 Heidelberg, Germany}

\received{...}
\accepted{...}
\publonline{later}


\abstract{%
Strong waves can mediate a shock transition between a pulsar wind and its
surroundings, playing the role of an extended precursor, in which the energy is
effectively transferred from fields to non-thermal particles.  The damping of such
precursors results in an essentially unmagnetized shock near the equator. 
In this context, we discuss the stability of strong waves and its implications
for the properties of shocks. Those with stable precursors
can exist in the winds of most of isolated pulsars, but the precursors may be unstable if
the external pressure in the nebula is high, as in Vela-like pulsars. Pulsar wind shocks in eccentric binary
systems, such as B1259$-$63, can acquire precursors only at certain orbital phases, and this process should be accompanied by enhanced
synchro-Compton and inverse Compton emission from the precursor. The same scenario may be at work in the binary 
HESS J0632+057.}

\maketitle

\newcommand{\eqb}{\begin{equation}}  
\newcommand{\eqe}{\end{equation}}			
\newcommand{\msolar}{\mbox{M$_\odot$}}
\newcommand{\lch}[1]{{#1}_0}
\newcommand{\pperp}{p_{\perp}}
\newcommand{\ppar}{p_{\parallel}}
\newcommand{\gammaw}{\gamma_{\rm w}}
\newcommand{\betaw}{\beta_{\rm w}}
\newcommand{\pext}{p_{\rm ext}}
\newcommand{\av}[1]{\left\langle #1 \right\rangle}
\newcommand{\derR}{\frac{1}{R^2}\frac{\partial}{\partial R}}
\newcommand{\derr}{\frac{1}{r^2}\frac{\partial}{\partial r}}
\newcommand{\der}[1]{\frac{\partial}{\partial #1}}
\newcommand{\bs}[1]{\mathbf{#1}}

\section{Introduction}

The presence of electromagnetic (EM) waves of large intensity close to highly magnetized stars, has been discussed since late 1960's.   
Before the discovery of pulsars \citet{1967Natur.216..567P}, and later also \citet{1969Natur.221..454G}, suggested that a spinning neutron star emits dipole radiation at the rotational frequency, 
which deposits fields, energy and radial momentum in the surrounding nebula.  
Moreover, the emission of a test particle accelerated in the high-intensity wave closely
resembles synchrotron emission in a static magnetic field, 
and \citet{1971ApJ...165..523G} attributed the continuum emission
from the Crab nebula to this so-called synchro-Compton (SC) process.
\citet{1971IAUS...46..407R} 
pointed out that
dipole radiation, and therefore also the SC emission, is linearly polarized at the equator, circularly polarized along the rotational
axis, and elliptically polarized at other latitudes. 
Since the linear component has the same direction at all latitudes, the model was in agreement with the observed uniform direction of polarization in the inner nebula. A contribution from the polar regions, however, was predicted to give a few percent of circular polarization, which was not found in the optical measurements of \citet{1971Natur.230..103L}.  
On the other hand, the observations
excluded only the model of vacuum dipole radiation. Additionally, the model of the pulsar magnetosphere
proposed by \citet{1969ApJ...157..869G} brought into consideration prolific pair production
and the existence of a plasma around a rotating neutron star. The need for a self-consistent
treatment of large-amplitude electromagnetic waves propagating in plasmas (\lq\lq strong waves\rq\rq) was realized \citep{1971PhRvL..27.1342M,1973PhFl...16.1277M,clemmow74}. These early
papers, however, showed that the propagation of such waves in the vicinity of pulsars is strongly
restricted: they cannot propagate in very dense media, and, therefore, one cannot expect them close to the light cylinder. They were also shown to be very unstable \citep{1972PhRvL..29.1731M,1978JPlPh..20..313L} and radiatively damped \citep{1978A&A....65..401A}. 

Pulsar spin-down power and angular momentum are thought, instead, to be carried away by a relativistic wind, a mixture of a plasma and frozen-in EM fields, launched close to the light cylinder $r_{\rm L}=c/\omega$. 
According to pulsar models, an assumed large pair production rate in the magnetosphere 
validates the magnetohydrodynamic (MHD) description of the wind. Far from the star, the dominant component of the magnetic field is toroidal, and the structure of the outflow is that of an entropy wave: it can be approximated by spherical current sheets, separating stripes of magnetized plasma with opposite magnetic polarity (\lq\lq striped wind\rq\rq). 
The wind is thought to power the diffuse emission of the nebula beyond the termination shock, 
which 
puts strict limits on the plasma magnetization downstream of the shock. 
However, according to theoretical pulsar models, the wind is highly magnetized when it is launched. Therefore, the question arises of how and where the wind dissipates its EM energy to the plasma. This so-called \lq\lq $\sigma$-problem\rq\rq\ cannot be solved by ideal MHD, because  
a radial highly magnetized MHD wind does not collimate as it propagates, and, therefore, does not convert the energy from the EM fields to kinetic form, and arrives at the shock still Poynting dominated. Shocks in magnetized flows are very weak, implying that 
the magnetization and the flow velocity practically do not change across them. 
The non-oscillating component of the fields in the wind has been shown to dissipate in the bulk of the nebula \citep{1998ApJ...493..291B,2013MNRAS.431L..48P}. Therefore the problem concerns mainly the dissipation of the wave-like oscillating component of the fields. \citet{2003MNRAS.345..153L} has proposed a solution, in which 
the striped wind, due to interaction with the shock, becomes compressed and the stripes dissipate the energy by the driven magnetic reconnection. Particle-In-Cell (PIC) simulations \citep{2011ApJ...741...39S} show that this mechanism operates also in 2D and 3D, but in order to reproduce the observed particle spectra, a very high plasma density has to be assumed. In fact, in plasmas of the assumed densities, reconnection in the wind would start much earlier, before it arrives at the shock \citep{2003ApJ...591..366K}. 

Thus, a complementary scenario must be at work when 
the plasma density is below the critical value. In a radial wind, the density drops with distance, and, beyond a certain point,  strong EM waves can propagate \citep{1975Ap&SS..32..375U,1996MNRAS.279.1168M}, both outwards from the star, and inwards from the shock.   
Because they have large intensities, they interact nonlinearly 
with each other. Instead of a  
superposition of outward propagating waves and reflected
waves, one must search for self-consistent solutions, containing both 
these components, and matching the outer boundary conditions. The simplest approach is to treat this process as a mode conversion between an MHD and a self-consistent EM wave, which forms a precursor to the shock. 
In this precursor a significant fraction of the flow energy is transported by the particles, which provides a solution to the $\sigma$-problem in the regime of low plasma density \citep{2010PPCF...52l4029K,2012ApJ...745..108A}. 
This approach is important to accurately describe the structure of shocks of isolated pulsars, as well as those in binary systems at certain orbital phases \citep{2013ApJ...771...53M}. In the latter case, one can expect a shock regime switch when it becomes possible for a shock to acquire a precursor. Close to this transition point, strong waves can be very efficient emitters, providing an explanation of the peculiar flare in the binary B1259$-$63 \citep{mocholkirk13}. 

In Sect.\ref{mhdwind} we introduce the MHD pulsar wind. In Sect.\ref{strongwaves} we describe the two-fluid model of EM waves and their propagation in pulsar winds, and in Sect.\ref{stability} we discuss their stability. The radiative signatures are presented in Sect.\ref{signatures}. The implications for the shocks of isolated PWNe, the binary system B1259$-$63, and a new prediction for the binary HESS J0632+057 are discussed in Sect.\ref{implications}.  

\section{The MHD pulsar wind}
\label{mhdwind}

The ability of a monochromatic, plane wave to accelerate a particle is quantified by the strength parameter: it expresses the Lorentz factor of a particle that it would gain if it were accelerated by the wave field of amplitude $E$ from rest over one wavelength $c/\omega$:
\eqb a=\frac{eE}{mc\omega} \eqe
We adapt this definition also to describe a general, nonvacuum wave. We assume that the wave carries the entire rotational power $L_{\rm sd}$ of the central compact object. 
In this case, the strength parameter at the light cylinder 
\eqb
a_{\rm L}=\left(\frac{4\pi e^2 L_{\rm sd}}{\Omega_{\rm s}m^2c^5}\right)^{1/2} 
= 3.4\times10^{10} L_{\rm sd,38}^{1/2}(4\pi/\Omega_{\rm s}) 
\eqe
where $\Omega_{\rm s}$ is the solid angle occupied by the wind. 
Note that since the luminosity carried by the wind decreases with the distance from the pulsar, the strength parameter also decreases $a=a_{\rm L}r_{\rm L}/r$. 

The particle flux density carried by the wind $J=\dot{N}/(r^2\Omega_{\rm s})$, where $\dot{N}$ is the pair production rate in the magnetosphere, also decreases with distance. Thus, the maximum available energy per particle in the wind is a distance-independent parameter
\eqb \mu=\frac{L_{\rm sd}}{\dot{N}mc^2}=\frac{a_{\rm L}}{4\kappa} \eqe
where $\kappa$ is the multiplicity (the ratio of the pair production rate to the Goldreich-Julian production rate), see \citet{2010PPCF...52l4029K} for a definition.  

The ratio of the Poynting flux to the particle energy flux carried by the wind defines the magnetization parameter $\sigma_0$. A cold, supermagnetosonic and magnetically dominated flow has 
\eqb 1\ll\sigma_0\lesssim \mu^{2/3} \eqe
The radial momentum of an MHD flow is another distance independent parameter: $\nu\approx\mu-\sigma_0/(2\mu)$. 
Its large value, $\nu\sim\mu$, implies that the particles move almost purely radially in the wind and the magnetization stays constant up to very large distances $r\sim a_{\rm L}r_{\rm L}/\sigma_0$, where charge-starvation forces a significant transverse momentum component and the magnetization starts decreasing. This is important in unconfined flows, like those of blazar jets \citep{2011ApJ...729..104K}. 

\section{Strong waves}
\label{strongwaves}
Nonlinear EM waves are exact solutions of the cold two-fluid ($e^{\pm}$) and Maxwell equations (plane waves are functions of only the phase variable). Their phase speed $\beta_{\rm ph}>1$ (superluminal modes), but the group speed is subluminal, $\betaw=1/\beta_{\rm ph}<1$. Large intensities ensure that the waves impart relativistic speeds on the particles within one period. Electrons and positrons have the same momenta $\ppar$ in the direction of the wave propagation; in the plane transverse to the wave propagation they have equal but oppositely directed momenta $\pm \bf{\pperp}$. In circularly polarized waves, the current that they generate exactly balances the displacement current of a wave $\left|\bf{\pperp}\right|=e \left|{\bf E}\right| / mc\omega$. 

The dispersion relation of circularly polarized electromagnetic waves in plasmas \citep[e.g.,][]{clemmow74}
\eqb \omega^2=\omega_{\rm p}^2+k^2c^2=\gammaw^2\omega_{\rm p}^2 \label{disprel} \eqe
where $\gammaw=(1-\betaw^2)^{-1/2}$, 
implies that they can propagate only when their frequency is larger than the local plasma frequency $\omega_{\rm p}=(8\pi e^2n/m)^{1/2}$ ($n$ is the proper density). In a pulsar wind two facts are important: (1) the frequency of an EM wave is fixed by the angular velocity of the neutron star $\omega$ and, therefore, it is convenient to measure $\betaw$ with respect to the pulsar frame in which the star is at rest; (2) according to the continuity equation, the local plasma frequency decreases with distance $\omega_{\rm p}^2\propto n\propto r^{-2}/\gamma$ (where $\gamma=(1+\pperp^2+\ppar^2)^{1/2}$ is the particle Lorentz factor in the pulsar frame). Thus, the wave propagation condition $\omega>\omega_{\rm p}$, following directly from Eq.~(\ref{disprel}), can be translated into a condition on the radius. With $\gamma\lesssim\mu$, one obtains  $r>r_{\rm crit}=a_{\rm L}r_{\rm L}/\mu$ and, therefore, a natural dimensionless radius is 
\eqb R=r/r_{\rm crit} \eqe
For the Crab, the critical radius $r_{\rm crit}\sim 10^6 r_{\rm L}$ is located well within the termination shock $r_{\rm ts}\sim 10^9r_{\rm L}$.

Mode conversion takes place at a somewhat larger distance $r_{\rm conv}>r_{\rm crit}$, uniquely determined for each pulsar by the boundary conditions (pressure) in the nebula beyond the shock. To determine the wave properties at the conversion point, one has to solve the electromagnetic jump conditions, assuming that both MHD and EM modes carry the same particle, energy and radial momentum fluxes \citep{2012ApJ...745..108A}. In the case of circular polarization the phase-averages can be dropped and, using definitions from Sect.\ref{mhdwind}, these can be expressed in the form
\begin{align}
J&=2c n\ppar 
\label{jumpconditions1}\\
\mu&=\left(\gamma+\frac{\betaw\gammaw^2\pperp^2}{\ppar}\right)
\label{jumpconditions2}\\
\nu&=\left(\ppar +\frac{\left(1+\betaw^2\right)\gammaw^2\pperp^2}{2\ppar}
\right)
\label{jumpconditions3}
\end{align}
In fact there are two possible solutions for an EM wave, whose fluxes equal those of the MHD wind: a free escape mode, and a confined mode. 
We concentrate on the latter, which at the conversion point has the following properties:
\eqb \gamma\approx\pperp\approx\mu/R_{\rm conv}, \quad \gammaw\approx(R_{\rm conv}/8)^{1/4} \eqe
(assuming $\pperp\gg\ppar$, valid everywhere except a very close neighbourhood of the cut-off point $R_{\rm conv}=1$). 

In the spherical expansion of the flow, the waveform is plane only to the lowest-order short-wavelength approximation. The first-order correction  describes the slow radial evolution of the lowest order, phase-averaged solution. For a circularly polarized wave the phase averages can be dropped and the evolution equations take the form:
\begin{align}
\frac{1}{r^2}\frac{\partial}{\partial r}\left(r^2n\ppar\right)&=0 \\ 
\frac{1}{r^2}\frac{\partial}{\partial r}\!\left(
r^22n\ppar
\gamma
+\frac{r^2\betaw
E^2}{4\pi mc^2}\right)
&=0 \\
\frac{1}{r^2}\frac{\partial}{\partial r}\left(
r^22n\ppar^2
+\frac{r^2\!\left(1+\betaw^2\right)E^2}{8\pi m c^2}\right)
&=\frac{n\pperp^2}{r}\enspace.
\label{momentum1}
\end{align}
The first equation immediately implies $\ppar=\mu\gammaw^2/R^2$. The other two are integrated with the initial conditions given by the jump conditions (\ref{jumpconditions2}-\ref{jumpconditions3}). Only the solution that starts at the confined mode branch  slows down with distance and eventually stagnates at a finite pressure. It can, therefore, be matched to the surroundings (see Fig.\ref{fig1}). We interpret it as an extended shock precursor.  

In a radially propagating wave the phase averaged Lorentz factor of the particles, measured in the laboratory frame, turns out to be conserved. This can be proven for arbitrary wave polarization \citep{2013ApJ...771...53M}. Thus, instead of equation (\ref{momentum1}) we can use
\eqb
\frac{\partial}{\partial r}\gamma=0\enspace.
\label{entropy1}
\eqe
This integral of motion is in fact an adiabatic invariant, and follows directly from the plasma neutrality in all reference frames related to the lab. frame by a Lorentz boost in the direction of wave propagation. 
It allows us to find the pressure at which a confined mode stagnates.

\section{The stability of strong waves}
\label{stability}

\begin{figure}
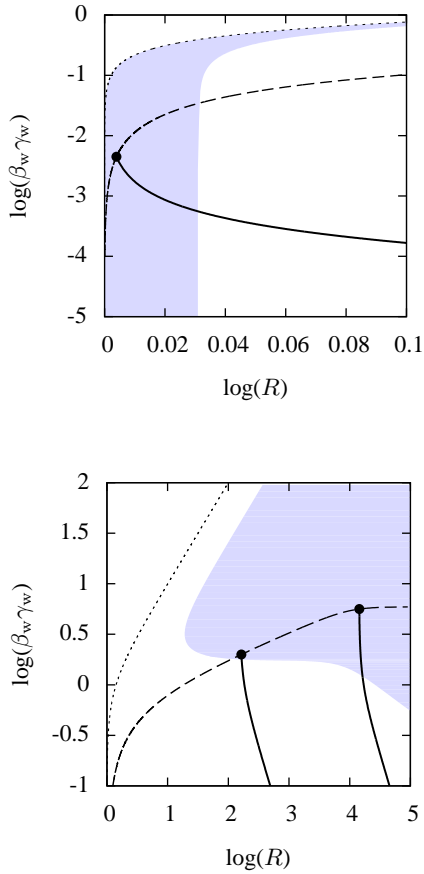

\centering
\input{fig1.tex}
\input{fig2.tex}
\caption{\label{fig1} 
Top panel: The inner zone of stability $1\!<\!R\!\lesssim\!(4/3)^{1/4}$. Electromagnetic Hugoniot curves are
plotted showing the group four-speed $\betaw\gammaw$
as a function of radius $R_{\rm conv}$ 
for $\mu =10^4$, $\sigma=100$ 
(dotted line: free expansion branch, dashed: confined branch). 
The black solid line shows the radial evolution of 
a wave, launched at the confined-mode branch. 
In the shaded region the waves are stable according to the 
criterion given in Eq.~(\ref{stcond}). Bottom: The outer zone of stability $R\gtrsim100$. The same Hugoniot curves
are plotted, together with two confined modes,
launched at larger radius. The inner zone of
stability lies close to the Hugoniot curve of the free escape mode,
and is not visible on the scale of this figure.
}
\end{figure}

Strong waves are intrinsically very unstable, unless they carry a transverse, phase-averaged component of the magnetic field \citep{1980PhRvA..22.1293A}, or there is a fast streaming of the particles through the wave \citep{1978JPlPh..20..313L,1978JPlPh..20..479R,2005ApJ...634..542S}. Here we consider only the second case as a stabilizing factor. A  stability condition for a circularly polarized wave has been obtained by \citet{1978JPlPh..20..313L}:
\eqb 
\ppar'^2-2\gamma'\pperp+\pperp^2\,>\,0\enspace
\label{stcond}
\eqe  
where a prime denotes the quantities measured in the frame comoving with the wave group speed. 
In pulsar winds relativistic streaming is a property of waves launched very close to the cut-off, $1< R\lesssim(4/3)^{1/4}$, or very far from it, $R\gtrsim100$ \citep{2013ApJ...771...53M}.  
The condition (\ref{stcond}) is shown as a shaded region in Fig.\ref{fig1}. 
The group four-speed of a wave $\gammaw\betaw$, calculated from the jump conditions (\ref{jumpconditions1}-\ref{jumpconditions3}), is shown as a
function of conversion radius $R_{\rm conv}$ (dotted line: free
expansion branch, dashed: confined branch). Solid curves show the radial
evolution of confined modes launched at different conversion radii (marked by a dot). Even if the wave is launched stable, it becomes unstable after propagation of $\sim10^3$ wavelengths. 
 
Where the wave is launched is uniquely determined by the external pressure  
\eqb p_{\rm ext}\approx\frac{L}{\Omega_{\rm s}c r^2}=\frac{10^{-6} \mu_4^2}{R^2 P_1}\enspace {\rm dyn\ cm}^{-2} \eqe 
where $P_1$ is the pulsar period in seconds, and $\mu=\mu_4\times10^4$. The waves launched in the outer stability zone match low pressure media like those of isolated pulsars, confined by nebulae at large distances. 
In the case of high-pressure nebulae the precursor is expected to fall into the unstable region. 
The waves launched in the inner stability zone match very high-pressure media like those of binaries, provided by the dense winds of their companion stars. However, in the absence of radiative damping this zone is very narrow.

\section{Radiative damping}
\label{signatures}

Waves launched close to the cut-off have the largest amplitudes, and, for them, radiation reaction can be an efficient mechanism of wave damping. The propagation of a wave in a planar geometry but with radiation reaction taken into account is governed by \citep{mocholkirk13}
\begin{align}
\frac{d}{dx}\left(n\ppar\right)&=0 \\
\frac{d}{dx}\left(2n\ppar\gamma+\frac{\betaw E^2}{4\pi mc^2}\right)&=ng^0  \label{energyrad} \\
\frac{d\gamma}{dx}&=\frac{g^1-\betaw g^0}{2\Delta}  \label{entropyrad}
\end{align}
where $\Delta=\gamma-\betaw\ppar$ and
\begin{align}
g^{0,1}&\approx-\frac{2e^4}{3m^3c^6}E^2\left|\gamma -\betaw\ppar\right|^2(\gamma,\ppar)\enspace.
\end{align} 

One can immediately see that the ratio of changes of $\gamma$ to changes of the Poynting flux per particle is $\sim\ppar^2/\pperp^2\ll1$. Therefore, the equations suggest that the particles catalyze extraction of the Poynting flux carried by the wave, i.e., the wave energy is used to accelerate particles by exactly the amount they lose to radiation, keeping the Lorentz factor constant. 

When in addition to radiation reaction IC scattering is also important, the radiative damping force acquires an additional  component, $g^{0,1}_{\rm ic}$ (see appendix \ref{append}). 

\begin{figure}
\centering
\input{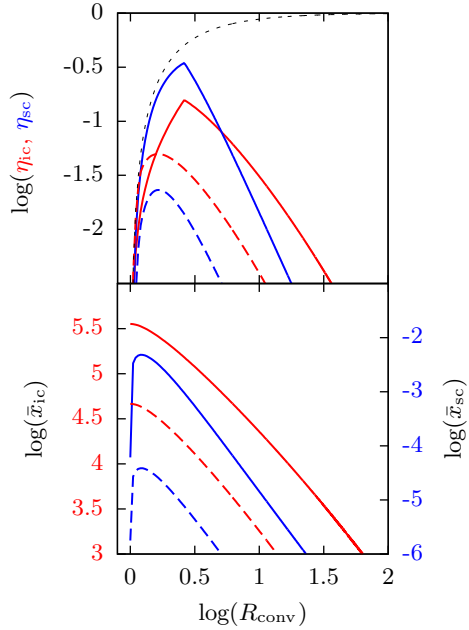}
\caption{\label{fig2} 
Radiative signatures of the shock precursor, for different conversion radii and for $a_{\rm L}=10^{10}$, and $\gamma_{\rm ic}/\gamma\approx3.9$. Red: IC component, blue: SC emission. Solid: $\mu=5\times10^5$, dashed $\mu=10^5$. Black dotted line shows the available Poynting flux. 
}
\end{figure}

Importantly, the radiative signatures, i.e., the efficiency of emission $\eta=\eta_{\rm sc}+\eta_{\rm ic}$, defined in Eq. (\ref{etatotal}), and the mean photon energy $\bar{x}=h\nu/mc^2$, defined in Eq. (\ref{xsc}) and (\ref{xic}) for SC and IC processes respectively, change with the distance of the shock (see Fig.\ref{fig2}). 
The overall emission efficiency in each component is an interplay between the amount of the Poynting flux available for extraction, the wave field strength (SC), and the energy density of the target photons (IC). 
In general, particles moving in a large-amplitude transverse wave, stay confined in the emission region longer than if they were moving purely radially.

\section{Implications}
\label{implications}

\subsection{PWNe}

PWNe can be divided into two groups: those with stable precursors to their termination shocks and those with unstable precursors. The first group includes Crab ($R=1355\kappa_{5}^{-1}$) and N158A ($R=1540\kappa_{5}^{-1}$). The second includes, for instance, Vela ($R=56\kappa_{5}^{-1}$), where $\kappa_5=\kappa/(5\times10^5)$.  
 
The simulations of \citet{2013ApJ...770...18A} have shown that an EM-modified shock behaves as essentially unmagnetized near the  equator. The reason is that the electric field $E>B$ dominates the plasma dynamics in the EM wave, and, therefore, particle transport does not have to rely on cross field diffusion, as in a perpendicular shock. In this sense, shock mediated by a stable precursor is expected to be stronger, and, therefore, a more efficient emitter. We speculate that the nebulae with stable/unstable precursors may exhibit different radiation signatures, for instance brighter/fainter appearance.   
However, the quantitative analysis is difficult, because nebular emission depends on many factors. 
In addition, pulsar multiplicities are unknown, although the  modelling of PWNe suggests large values $\kappa\gtrsim{\rm a\ few}\times10^5$ \citep{2011MNRAS.410..381B}. 

\subsection{The binary B1259$-$63} 
An interesting possibility arises in binary systems, where the shocks may switch between different regimes. When the binary members are close, such that the distance between the pulsar and the shock is smaller than the critical radius, the shock is in the MHD regime. When the separation becomes larger, a shock can acquire an EM precursor. A sudden appearance of a precursor ahead of the shock should be accompanied by enhanced emission in the SC and IC processes (see Fig.\ref{fig2}) close to this  transition point, because the precursor emission is most efficient close to the cut-off. 

An example is the eccentric binary B1259$-$63. It consists of a pulsar on a very elongated orbit around a Be-star, with a period 3.4 years; 30 days after the periastron passage in 2010, Fermi-LAT detected a GeV flare that lasted several weeks \citep{2011ApJ...736L..10T,2011ApJ...736L..11A}. The efficiency of this emission was extremely high, on the order of the pulsar spin-down power. This feature is challenging for all existing models: those based on electrons accelerated at the termination shock
\citep{2012ApJ...753..127K} have to assume strong Doppler-beaming, those based on electrons in the unshocked wind
\citep{2011MNRAS.417..532P,2011ApJ...742...98K,2012ApJ...752L..17K}  
require an additional source of target photons.  

The rotation period 48-ms of a pulsar implies $r_{\rm L}=2.3\times10^8$ cm, and spin down power $L_{\rm sd}=8\times10^{35}$ erg s$^{-1}$ is equivalent to the strength parameter $a_{\rm L}=3\times10^9 (4\pi/\Omega_{\rm s})$. 30 days after periastron passage the separation between binary members is $3.7\times10^{13}$ cm, and we assume that the shock is located roughly a mid-way between the objects. Since the flare occurs close to $R_{\rm conv}\approx1.5$, we obtain $\mu\sim 6\times10^4$ and the characteristic energy of IC photons $\epsilon_0\sim 10$ GeV, which decreases with time.  

We also expect a faint (${\rm a\ few}\times 10^{-3} L_{\rm sd}$) counterpart due to synchro-Compton emission in the optical band. However, this component is swamped by photons from the luminous companion star. 
 
In principle, an analogous, pre-periastron flare could be expected 
if the physical conditions at the shock were symmetric along the pulsar orbit with respect to periastron. These conditions are determined by the location of a shock between the pulsar and stellar winds, i.e., by the wind relative strength. Since the wind of the companion star is known to be highly anisotropic, the conditions at the shock are unlikely to be symmetric with respect to periastron. Therefore, a pre-periastron flare  
is not necessarily expected to be symmetrically timed with the post-periastron flare, and, consequently, not necessarily peaked in the same energy band.  

\subsection{Predictions for the binary HESS J0632+057}
The recently discovered gamma-ray binary HESS J0632+057 consists of an unknown compact object on an elongated orbit (period 315 days) around a Be-star. The lightcurve exhibits enhancement of X-ray and sub-TeV emission 100 days after the periastron, and located roughly symmetrically with respect to the apastron passage \citep{2012AIPC.1505..366B}. If a binary member were a pulsar, one could expect similar behaviour as in B1259$-$63. 
The sub-TeV emission due to IC scattering of the stellar photons implies $\mu\gtrsim3.5\times10^5$, and the SC emission in keV suggests the pulsar period $\lesssim 140$~ms. The maximum emission 100 days after periastron implies $a_{\rm L}\approx9\times10^9$.

Thus, our estimates show that the model can account for simultaneous emission in both energy bands with their unusual lightcurves (two maxima and a dip around the apastron), and predict the presence of a pulsar with period $P\lesssim140$~ms and period derivative $\dot{P}\approx4.4\times10^{-13}$.   

\section{Summary and conclusions}

The dynamics of pulsar winds is dominated by electromagnetic fields, which, far from the star, behave more like an EM wave instead of a familiar MHD wind. These effects are crucial to a proper description of shocks in a low-density plasma, i.e., those  located at large stand-off distances, $r_{\rm ts}\gtrsim10^2r_{\rm crit}$. 
Such shocks exhibit dissipative precursors, which modify their structure and particle acceleration properties. In most of the isolated pulsar winds, the precursors are stable over many wavelengths, but if the surrounding nebula is compact, like Vela, the precursors are unstable. In eccentric binaries it is possible to probe a shock regime switch between an MHD shock and an electromagnetically modified shock, and the apperance of a precursor shock can be distinguished by its radiative signatures. 

\acknowledgements
IM thanks Vincent Marandon for useful discussions.    




\appendix

\section{Efficiency of emission}
\label{append}
We introduce the space-dependent magnetization parameter
\begin{align}
\sigma&=\frac{ \betaw \pperp^2}{(8\pi e^2 n/m\omega^2) \ppar\gamma}\\
&=\frac{ \gammaw^2\betaw \pperp^2}{\ppar\gamma}
\end{align}
where we have used the dispersion relation for strong waves $8\pi e^2 n/m\omega^2=1/\gammaw^2$ \citep{clemmow74}. 
This parameter allows us to express the Poynting flux per particle by $\gamma\sigma$, and the total energy per particle $\mu=\gamma(1+\sigma)$.  
Eq. (\ref{energyrad}) and (\ref{entropyrad}) imply that the Poynting flux decreases as 
\eqb \frac{d}{dX}\left( \gamma\sigma \right)=-\frac{\epsilon a_{\rm L}}{2\mu}\frac{\Delta \pperp^2 (1+\pperp^2)}{\ppar} \eqe
where $X=x/r_{\rm crit}=(\mu/a_{\rm L})(x/r_{\rm L})$ and $\epsilon=2e^2\omega/3mc^3$. 

One can estimate a lengthscale, on which a significant fraction of the Poynting flux is extracted
\eqb X_{\rm diss}\approx \frac{2\mu}{\epsilon a_{\rm L}}\left(\frac{\gamma\sigma\ppar}{\Delta\pperp^2\left(1+\pperp^2\right)}\right) \eqe
When the IC emission is important, the radiative coefficient is two-component 
\begin{align}
\epsilon&=\epsilon_{\rm sc}+\epsilon_{\rm ic} \\
&=\frac{2e^2\omega}{3mc^3}\left(1+\frac{G(\gamma,x_0)\gamma_{\rm ic}^2}{\gamma^2}\right) \\
\noalign{where}
\gamma_{\rm ic}&=\left(\frac{2\sigma_{\rm T}U_{\rm rad} c^2}{e^2\omega^2}\right)^{1/2}
\end{align}
and $G(\gamma,x_0)$ is the reduction factor due to Klein-Nishina effects in the scattering of a target photon field $\nu_0=x_0 mc^2/h$ \citep[see, e.g.,][]{1999APh....10...31K}. 
A substantial fraction of the Poynting flux is converted into radiation if $X_{\rm diss}<R_{\rm conv}$, hence the radiation efficiency can be defined as
\begin{align}
\eta&=
\left\lbrace
\begin{array}{ll}
\sigma/(1+\sigma)  & \textrm{for\ }X_{\rm diss}<R_{\rm conv}\\
\sigma R_{\rm conv}/\left[X_{\rm diss}(1+\sigma)\right]  & \textrm{for\ }X_{\rm diss}>R_{\rm conv}
\label{etatotal}
\end{array}
\right.
\end{align}
The mean energy of radiated photons in the synchro-Compton process 
\eqb \bar{x}_{\rm sc}(\gamma,\pperp)=0.4\gamma^2\pperp \left(\hbar\omega/mc^2\right), \label {xsc}\eqe 
whereas the peak energy of inverse Compton scattered photons can
be estimated as
\eqb
\bar{x}_{\rm ic}(\gamma,x_0)=(4/3)\sigma_{\rm T}x_0 \gamma^2 G(\gamma,x_0)
/\langle \dot{N}_\gamma\rangle, \label{xic}
\eqe
where $\dot{N}_\gamma$ is the scattering rate divided by the density of
target photons and $\langle\dots\rangle$ indicates an angle average, 
see \cite{mocholkirk13}.

\end{document}